\documentstyle[preprint,aps,epsf,cite]{revtex}

\newcommand{\newc}{\newcommand}
\newc{\beq}    {\begin{equation}}
\newc{\eeq}    {\end{equation}}
\newc{\beqa}    {\begin{eqnarray}}
\newc{\eeqa}    {\end{eqnarray}}
\newc{\no}    {\\ \nonumber}

\begin{document}
\draft


\title{Nonlocality of Hardy type in
experiments using independent particle sources}

\author{Xu-Bo Zou \footnote{Present address: Electromagnetic Theory Group at THT,
 Department of Electrical Engineering,
University of Hannover, Germany}
and  Hai-Woong Lee}
\address{
Department of Physics, Korea Advanced Institute of Science and
Technology, Taejon 305-701, Korea \\}
\author{ Jaewan Kim }
\address{
School of Computational Sciences,
Korea Institute for Advanced Study,
207-43 Cheongryangri-dong, Dongdaemun-gu, Seoul 130-012, Korea}

\author{ Jae-Weon Lee and  Eok Kyun Lee  \\}
\address{
 Department of Chemistry,  School of Molecular Science (BK 21),
Korea Advanced   Institute of Science and Technology,  Taejon
 305-701, Korea.}

\maketitle

\begin{abstract}
{\normalsize By applying Hardy's argument, we demonstrate the
violation of local realism in a gedanken experiment using
independent and separated particle sources. }
\end{abstract}

The nonlocal nature of quantum systems  arising
from entanglement  has played a central role in quantum information
science.
Discussions about quantum  nonlocality were initiated by
Einstein, Podolsky, and Rosen\cite{epr} and extended by
Bell\cite{bell}.
Although the violation of Bell's inequality predicted
by quantum mechanics
 has been experimentally verified\cite{exp},
there have been arguments about the detection
loopholes\cite{loophole,hwang}.
Greenberger, Horne,
and Zeilinger(GHZ) \cite{ghz} demonstrated
 quantum mechanical violation of local realism
without using the Bell's inequality for more than three
particles. Hardy proved the nonlocality without using the Bell's
inequality
  for all entangled states (except maximally entangled
states) of two spin-$\frac{1}{2}$ particles\cite{hardy}.
 Considerable theoretical and
experimental effort has been devoted to testing this Hardy type
nonlocality\cite{goldstein,hardy2,hwang,wu,yurke}. An attempt to
extend Hardy's theorem to cover maximally entangled states was
made by Wu et al. using a quantum optical setting\cite{wu}.
Recently Yurke and Stoler demonstrated violation of local realism
in an experimental configuration involving independent
sources\cite{yurke,yurke2,yurke3}. Specifically, they showed that:
(1) in the fermion case the Pauli exclusion principle can be
exploited in a local realism experiment of the Hardy
type\cite{yurke}
 ; (2)  GHZ type nonlocality can arise even when
the particles come from independent widely separated
sources\cite{yurke2}; and (3) violation of the Bell's inequality
can be demonstrated by a quantum optical setting  using
independent particle sources \cite{yurke3}. \\
\indent
 The aim of our paper is to demonstrate
nonlocality of Hardy type in experiments using independent
particle sources. A schematic of the apparatus for the Gedanken
experiment is shown in Fig. 1, which is similar to the setup
proposed by Yurke and Stoler \cite{yurke3}, except that the four
beam splitters $B_i$ have transmittance $T_i$ and reflectivity
$R_i=1-T_i$ where $T_i \neq R_i$ and $i=1..4$. We parameterize
$T_i$ and $R_i$ as ${T_i}=sin^2(\theta_i)\equiv S^2_i$ and
${R_i}=cos^2(\theta_i)\equiv C^2_i$, respectively.
\begin{figure}[Fig1]
\epsfysize=6cm \epsfbox{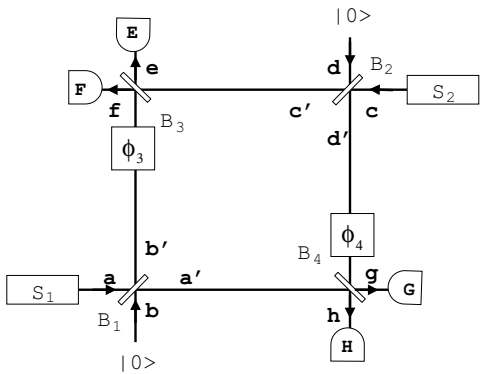} \caption[Fig1]{\label{fig1}
Schematic of the  apparatus used in our gedanken experiment for
Hardy's nonlocality. See text for detailed explanations. }
\end{figure}
 Two independent particles
radiated from the source $ S_1$ and $S_2$ are incident on the
input ports of the beam splitters $B_1$ and $B_2$, respectively.
 Vacuum ($|0\rangle$) enters the other input ports of
$B_1$ and $B_2$. The outputs of these beam splitters propagate to
two detectors. Detector 1 consists of the phase shifter $\phi_3$,
the beam splitter $B_3$, and the particle counters E and F.
Similarly, detector 2 consists of the phase shifter $\phi_4$, the
beam splitter $B_4$, and the particle counters G and H. The beam
path labels appearing in Fig. 1 will also be used to denote the
annihilation operators for modes propagating along these  beam
paths.

\indent The analysis of the firing statistics at each particle
counter is
 carried out as follows.
The beam splitters $B_1$, $B_2$, $B_3$, and $B_4$
perform the mode transformation
\beqa
\label{trans}
  \left (
         \begin{array}{cc}
          a'\\
          b'
           \end{array}
    \right )&=&\left (
         \begin{array}{cc}
          S_{1}&iC_{1}\\
          iC_{1}&{S_1}
           \end{array}
    \right )
    \left (
         \begin{array}{cc}
          a\\
          b
           \end{array}
    \right ),
    \no
  \left (
         \begin{array}{cc}
          c'\\
          d'
           \end{array}
    \right )&=&\left (
         \begin{array}{cc}
          S_{2}&iC_{2}\\
          iC_{2}&S_{2}
           \end{array}
    \right )\left (
         \begin{array}{cc}
          c\\
          d
           \end{array}
    \right ),
    \no
  \left (
         \begin{array}{cc}
          e\\
          f
           \end{array}
    \right )&=&\left (
         \begin{array}{cc}
          S_{3}&iC_{3}\\
          iC_{3}&S_{3}
           \end{array}
    \right )\left (
         \begin{array}{cc}

           e^{-i{\phi_3}}b'\\
          c'
           \end{array}
    \right ),
    \no
  \left (
         \begin{array}{cc}
          g\\
          h
           \end{array}
    \right )&=&\left (
         \begin{array}{cc}
          S_{4}&iC_{4}\\
          iC_{4}&S_{4}
           \end{array}
    \right )\left (
         \begin{array}{cc}
          a'\\
          e^{-i\phi_4}d'
           \end{array}
    \right ).
\eeqa
 From the mode transformation shown in Eq. (1),
it follows that  the annihilation operators for the modes
$a,b,c$ and $d$  can be expressed in terms of those
 for  $e,f,g$ and $h$ as follows:
\beqa
 \label{a}
a&=&S_1 S_4g-iS_1 C_4h
 -  ie^{i\phi_3}{C_1 S_3}e
 -e^{i\phi_3}{C_1 C_3}f \no
b&=&-i{C_1 S_4}g-{C_1
 C_4}h
 + e^{i\phi_3}{S_1 S_3}e
 -ie^{i\phi_3}{S_1 C_3}f \no
 c&=&-i{S_2
C_3}e+{S_2 S_3}f  -e^{i\phi_4}{C_2 C_4}g
-ie^{i\phi_4}{C_2 S_4}h \no
 d&=&-{C_2
C_3}e-i{C_2 S_3}f  -ie^{i\phi_4}{S_2 C_4}g
+e^{i\phi_4}{S_2 S_4}h.
\eeqa
The state vector for
 two identical bosons  injected into the interferometer
can be expressed as the direct product of the individual state
vectors. In  second quantized notation, the input state vector is
therefore given by \beq \label{psi}
|\psi\rangle=a^\dagger c^\dagger |0\rangle. \eeq By substituting
Eq. (\ref{a}) into Eq. (\ref{psi}), we obtain the output state
vector. This vector can be divided into two parts \beq
|\psi\rangle=|\psi_1\rangle+|\psi_2\rangle, \eeq where \beqa
 |\psi_1\rangle&=& i[{S_1 S_2 C_3 S_4}
-e^{-i\phi_3-i\phi_4}{C_1 C_2 S_3
C_4}]|1\rangle_e|1\rangle_g \no
 &+&[{S_1 S_2 S_3 S_4}
+e^{-i\phi_3-i\phi_4} {C_1 C_2 C_3
C_4}]|1\rangle_f|1\rangle_g \no
 &-&[{S_1 S_2 C_3 C_4}
+e^{-i\phi_3-i\phi_4}{C_1 C_2 S_3
S_4}]|1\rangle_e|1\rangle_h \no
 &+&i[{S_1 S_2 S_3 C_4}
-e^{-i\phi_3-i\phi_4}{C_1 C_2 C_3
S_4}]|1\rangle_f|1\rangle_h, \no
|\psi_2\rangle&=&ie^{-i\phi_3}{C_1 S_2}(S_3^2 -C_3^2)
|1\rangle_e|1\rangle_f \no
&+&ie^{-i\phi_4}{S_1 C_2}(S_4^2-C_4^2)|1\rangle_g|1\rangle_h \no
&-&{2}e^{-i\phi_3} {C_1
S_2S_3C_3}[|2\rangle_e+|2\rangle_f] \no
&-&{2}e^{-i\phi_4}
{S_1C_2S_4C_4}[|2\rangle_g+|2\rangle_h].
\eeqa Here $|n\rangle_e$ denotes the $n$ particle state of the
mode $e$. Now consider detector 1 and let $\bar{E}~$($\bar{F}$)
denote the event in which the counter E (F) counts  a single
particle  and the counter
 F (E)  counts no particle.
 Similarly for detector 2, we define events
$\bar{G}~(\bar{H})$ in which the counter
 G (H) counts one particle and the counter H (G)
counts no particle.
Among all the possible events, we are interested only in
the events
$\bar{E},\bar{F},\bar{G}$ and $\bar{H}$.
 Thus, we do not need to pay
  attention
to the evolution of the state $|\psi_2\rangle$, because
$\bar{E},\bar{F},\bar{G}$ and $\bar{H}$ are not reflected in
  $|\psi_2\rangle$. Hence, below we consider only the evolution
of the state $|\psi_1\rangle$.

Let us consider the  following
four cases.\\
a) Set $\phi_3 =\phi_4=\pi/2$ and choose
\beq
S_3=S_4=\sqrt{\frac{{C_1C_2}}{{C_1C_2}+{S_1S_2}}}
\equiv \tau.
\eeq
Then we obtain
\beqa
\label{psi1}
|\psi_1\rangle&=& i\sqrt{C_1C_2S_1S_2}
(|1\rangle_e|1\rangle_g +|1\rangle_f|1\rangle_h) \no
 &+&[{C_1C_2} -{S_1S_2}]|1\rangle_e|1\rangle_h.
\eeqa
 Let
$FG(S_3=\tau,S_4=\tau,\phi_3=\frac{\pi}{2},\phi_4=\frac{\pi}{2})$
denote the probability of the simultaneous
appearance of events $\bar{F}$
 and $\bar{G}$
for the following experimental settings.
The transmittance  of $B_3$
is set to $S_3^2=\tau^2$ and that of $B_4$ is set to
$S^2_4=\tau^2$.
 Since there is
no $|1\rangle_f|1\rangle_g$ term in Eq. (\ref{psi1}), we obtain
\beq
FG(S_3=\tau,S_4=\tau,\phi_3=\frac{\pi}{2},\phi_4=\frac{\pi}{2})=0.
\label{fg1}
\eeq
\\
b) If $\phi_3=\frac{\pi}{2}$,
$\phi_4=\frac{3\pi}{2}$, and
\beqa
S_3&=&\tau, \no
S_4&=&\sqrt{\frac{(C_1C_2)^3}{(S_1S_2)^3 +
(C_1C_2)^3}}\equiv \tau',
\label{r4t4}
\eeqa
we have
\beqa
|\psi_1\rangle&=&
 \frac{{S_1S_2C_1C_2}}
{\sqrt{(C_1C_2)^2 +(S_1S_2)^2-{S_1S_2C_1C_2}}}
 |1\rangle_f|1\rangle_g \no
&-&\sqrt{(C_1C_2)^2+(S_1S_2)^2-{S_1S_2C_1C_2}} |1\rangle_e|1\rangle_h \no
&+&i\frac{\sqrt{S_1S_2C_1C_2}({S_1S_2}-{C_1C_2})}
{\sqrt{(C_1C_2)^2+(S_1S_2)^2-{S_1S_2C_1C_2}}}
|1\rangle_f|1\rangle_h.
\eeqa
Thus we have the following quantum prediction:\\
$F(S_3=\tau,S_4=\tau',\phi_3=\frac{\pi}{2},
\phi_4=\frac{3\pi}{2})=1,$
if
\beq
\label{gf1}
G(S_3=\tau,S_4=\tau',\phi_3=\frac{\pi}{2},
\phi_4=\frac{3\pi}{2})=1,
\eeq
since there is only one term $|1\rangle_f|1\rangle_g$
containing $|1\rangle_g$.\\
\\
c)  Setting $\phi_3=\frac{3\pi}{2}$, $\phi_4=\frac{\pi}{2}$ and
\beqa \label{r3t3} S_3&=&\tau', \no S_4&=&\tau, \eeqa we obtain
\beqa
  |\psi_1\rangle&=&
  \frac{{S_1S_2C_1C_2}}
{\sqrt{(C_1C_2)^2+(S_1S_2)^2-{S_1S_2C_1C_2}}}
 (|1\rangle_f|1\rangle_g) \no
&-& \sqrt{(S_1S_2)^2 + (C_1C_2)^2 -{S_1S_2C_1C_2}}
{ (|1\rangle_e|1\rangle_h) }\no
&+&i \frac{ \sqrt{S_1S_2C_1C_2}({S_1S_2}-{C_1C_2})}
{\sqrt{(C_1C_2)^2 + (S_1S_2)^2-{S_1S_2C_1C_2}}}
{ (|1\rangle_e|1\rangle_g)}.
\eeqa
Thus, if
 $$F(S_3=\tau',S_4=\tau
 ,\phi_3=\frac{3\pi}{2},\phi_4=\frac{\pi}{2})=1,$$
then
\beq
G(S_3=\tau',S_4=\tau,\phi_3=\frac{3\pi}{2},
\phi_4=\frac{\pi}{2})=1.
\label{fg2}
\eeq
\\
d) Setting $\phi_3=3\pi/2, \phi_4=3\pi/2$,
and choosing $S_3=S_4=\tau'$,
we obtain
\beqa
  |\psi_1\rangle&=&
  i\frac{\sqrt{(C_1C_2S_1S_2)^3}}
{(S_1S_2)^2+(C_1C_2)^2-{S_1S_2C_1C_2}}
 (|1\rangle_e|1\rangle_g \no
 &+& |1\rangle_f|1\rangle_h )
+ \frac{ {S_1S_2C_1C_2}({C_1C_2}-{S_1S_2})} {(S_1S_2)^2 +
(C_1C_2)^2-{S_1S_2C_1C_2}} { (|1\rangle_f |1\rangle_g)}\no &+&
\frac{ [(C_1C_2)^2+(S_1S_2)^2]({C_1C_2}-{S_1S_2})}
{(S_1S_2)^2+(C_1C_2)^2-{S_1S_2C_1C_2}} { (|1\rangle_e
|1\rangle_h)}. \eeqa Thus  the following quantum prediction is
obtained \beq
FG(S_3=\tau',S_4=\tau',\phi_3=\frac{3\pi}{2},\phi_4=\frac{3\pi}{2})=1
\label{fg3} \eeq with a nonzero probability \beq P=\frac{
(S_1S_2C_1C_2)^2({S_1S_2}-{C_1C_2})^2} {[(S_1S_2)^2+(C_1C_2)^2 -
{S_1S_2C_1C_2}]^2}, \eeq where $C_1C_2 \neq S_1S_2$. \\
\indent Finally, we demonstrate that, following
Hardy's\cite{hardy} and Wu et al.'s argument\cite{wu}, local
realism and quantum mechanics are incompatible using an
experimental setting with independent and separated particle
sources. The notion of local realism is introduced by assuming
that there exist some hidden variables $\lambda$ that describe
the state of individual particles. According to the assumption of
locality,
 the choice of the measurement at detector 1 would not
influence the outcome of the measurement at detector 2, which
means that, for a specified $\lambda$ ,
 the probability of the event
$\bar{F}$ is uniquely determined by
 the transmittance of $B_3$ and $\phi_3$, whereas
that of  $\bar{G}$ is determined solely by the transmittance of
$B_4$ and $\phi_4$.
Let us denote the probabilities of the
events $\bar{F}$  and
 $\bar{G}$ for a value of  hidden variable
 $\lambda$ by $F(\lambda,S_3,\phi_3)$  and
$G(\lambda,S_4,\phi_4)$, respectively.
 Using Eq. (\ref{fg3}), for some values of
hidden variable $\lambda$,
we expect  simultaneous occurrence of
events $\bar{F}$ and  $\bar{G}$ when
$S_3=\tau',S_4=\tau',\phi_3=\frac{3\pi}{2}$ and $\phi_4=\frac{3\pi}{2}$
and thus obtain
$F(\lambda,S_3=\tau',\phi_3=\frac{3\pi}{2})
=G(\lambda,S_4=\tau',\phi_4=\frac{3\pi}{2})=1$.
 On the other hand, from Eqs.
(\ref{gf1}) and (\ref{fg2}), we have
$G(\lambda,S_4=\tau,\phi_4=\frac{\pi}{2})=1$, since
$F(\lambda,S_3=\tau',\phi_3=\frac{3\pi}{2})=1$; and
$F(\lambda,S_3=\tau,\phi_3=\frac{\pi}{2})=1$ since
$G(\lambda,S_4=\tau',\phi_4=\frac{3\pi}{2})=1$
for the same values of $\lambda$. Therefore, we
should have $F(\lambda,S_3=\tau,\phi_3=\frac{\pi}{2})
=G(\lambda,S_4=\tau,\phi_4=\frac{\pi}{2})=1$.
But this  contradicts
the quantum prediction of Eq. (8) that
the probability of simultaneous occurrence of the events
$\bar{F}$ and $\bar{G}$ is zero when
$S_3=\tau,S_4=\tau,\phi_3=\frac{\pi}{2}$ and $\phi_4=\frac{\pi}{2}$.\\
\indent In summary, we have shown the violation of local
realism of EPR type without using Bell's inequality
for the case of  two
particles originating  from independent
sources.
\newline
\\
 We acknowledge the
support of the Brain Korea 21 Project of the Korean Ministry of
Education.

\end{document}